\documentclass[manuscript,screen]{acmart}
\makeatletter
\def\@ACM@checkaffil{
    \if@ACM@instpresent\else
    \ClassWarningNoLine{\@classname}{No institution present for an affiliation}%
    \fi
    \if@ACM@citypresent\else
    \ClassWarningNoLine{\@classname}{No city present for an affiliation}%
    \fi
    \if@ACM@countrypresent\else
        \ClassWarningNoLine{\@classname}{No country present for an affiliation}%
    \fi
}
\makeatother
\AtBeginDocument{%
  \providecommand\BibTeX{{%
    \normalfont B\kern-0.5em{\scshape i\kern-0.25em b}\kern-0.8em\TeX}}}

\setcopyright{acmlicensed}
\copyrightyear{2024}
\acmYear{2024}
\acmDOI{}

\usepackage[utf8]{inputenc}  

\usepackage{CJKutf8} 
\begin{document}
\begin{CJK}{UTF8}{gbsn}
\title{Teaching Linguistic Justice through Augmented Reality}


\author{Ashvini Varatharaj}
\email{ashvinivaratharaj@ucsb.edu}
\orcid{0000-0001-6714-0743}
\affiliation{%
\institution{University of California, Santa Barbara}
  \city{Santa Barbara}
  \state{California}
  \country{USA}}

\author{Abigail Welch}
\email{awelch@ucsb.edu}
\orcid{}
\affiliation{%
\institution{University of California, Santa Barbara}
  \city{Santa Barbara}
  \state{California}
  \country{USA}
  }

\author{Mary Bucholtz}
\email{bucholtz@ucsb.edu}
\orcid{0000-0001-7343-3374}
\affiliation{%
\institution{University of California, Santa Barbara}
  \city{Santa Barbara}
  \state{California}
  \country{USA}
}

\author{Jin Sook Lee}
\email{jinsooklee@ucsb.edu}
\orcid{0000-0001-9812-4922}
\affiliation{%
\institution{University of California, Santa Barbara}
  \city{Santa Barbara}
  \state{California}
  \country{USA}
}

\renewcommand{\shortauthors}{Varatharaj, et al.}

\begin{abstract}
 This position paper presents the AR Language Map, a speculative artifact designed to enhance understanding of linguistic justice among middle and high school students through augmented reality (AR) that allows students to map their linguistic experiences. Through a social justice-oriented academic outreach program aimed at linguistically, economically, and racially minoritized students, academic concepts on language, culture, race, and power are introduced to California middle school and high school students. The curriculum has activities for each lesson plan drawn from students’ culturally relevant experiences. By enabling interactive exploration of linguistic justice, this tool aims to foster empathy, challenge linguistic racism, and valorize linguistic diversity. We discuss its conceptualization within the broader context of AR in social justice education. The AR Language Map not only deepens students' understanding of these critical issues but also enables them to become co-creators of their learning experiences.

\end{abstract}






\maketitle

\section{Introduction}
Linguistic justice is a concept promoted by growing numbers of linguists who advocate for respect and recognition of individual and community language rights, emphasizing the importance of communication in one's preferred language(s) \cite{avineri2019language}. This concept is intrinsically connected to broader justice issues, such as racial and educational justice, because language is both a marker of identity and a basis for discrimination and exclusion. Teaching linguistic justice is vital not only for fostering an inclusive and empathetic society but also for supporting students’ agency to express themselves in their own languages \cite{baker2017can}. By integrating linguistic justice into educational curricula, we can challenge and help to dismantle systemic inequities perpetuated through linguistic racism and other forms of linguistic oppression. Using AR as an innovative tool for language mapping, can promote critical consciousness about how students and others navigate language through immersion in the world of their ‘language maps’.

\section{Related works}
\subsection{Education towards teaching linguistic justice}
Education is imbued with power dynamics around language, which is often used as a gatekeeping mechanism for students from minoritized backgrounds whose languages are traditionally viewed as ‘inferior’ to prescriptivist ‘standard’ English. This creates and perpetuates educational inequalities and affects student performance. Moreover, global flows, immigration and transnational mobility brings new languages and linguistic varieties into schools, yet newcomer students’ linguistic backgrounds are often poorly supported (e.g., \cite{perez2016zapotec}). Educational initiatives have increasingly recognized the importance of incorporating linguistic justice into curricula. Researchers have argued that the benefits of decades of linguistic research in schools have not reached students and that future research and applications must address this gap. When minoritized students’ language experiences are valued, they start to utilize their linguistic capital as a resource and recognize their linguistic agency, which helps them to thrive in the educational system \cite{mallinson2024linguistic}. Educators have a crucial role to play by helping students develop critical linguistic awareness and combat societal-level language ideologies. In addition, research-based educational programs that work in partnership with linguistically subordinated students and their teachers are necessary to challenge dominant language ideologies and bring linguistic justice into classrooms \cite{avineri2019language, bucholtz2014sociolinguistic}.

\subsection{Experiences of Latinx children of immigrants}
In California, most students in public schools are Latinx children of immigrants and are often immigrants themselves. Prior research on Chicanx and Latinx students has brought attention to how their language experiences shape their identity and tap into their ‘funds of knowledge’ by utilizing the skillsets they have already started building at home \cite{moll2006funds}. These studies advocate for an educational approach that transcends traditional, restrictive views of language learning and teaching, highlighting instead the rich, dynamic linguistic capacities that Chicanx and Latinx students bring to the classroom. Our proposal seeks to promote a translanguaging pedagogical approach that embraces all language resources each student has at their disposal \cite{garcia2015translanguaging}.

\subsection{Linguistic Cartography}
An important method for student exploration of their own and others’ language backgrounds and experiences is language mapping \cite{martinez2020looking}, or linguistic cartography, which guides students to consider how different spaces influence their language practices and vice versa. Prior work has shown that linguistic cartography is a powerful tool to help students reflect on all of their language resources learned across a range of contexts and social purposes \cite{phillips2022linguistic}. Linguistic cartography also provides a platform for questioning and challenging educational and societal discourses that pathologize multilingualism. It can be a unique tool for documenting how students’ linguistic identities are shaped and changed by “the spaces, bodies, and materials we inhabit” \cite{lytra2022liberating}. This perspective aligns with the objectives of our project, which seeks to harness AR technology to create immersive, interactive learning experiences that acknowledge and celebrate students' multifaceted linguistic identities.

\subsection{AR as a Tool for Pedagogical, Social and Linguistic Justice }
Augmented Reality (AR) has numerous demonstrated benefits in education, such as enhancing long-term memory retention, fostering increased collaboration, and facilitating the creation of interactive digital spaces for students to work together and construct collective understanding \cite{radu2014augmented, morrison2009like}. However, AR’s potential to promote educational equity remains underutilized, often due to a mismatch between technological applications and the students' interests or a lack of emphasis on student agency. Within linguistics, the bulk of existing research on AR focuses on language acquisition \cite{zhang2018augmented}, our initiative seeks to leverage AR to explore and impart concepts of linguistic justice, the first use of AR for this purpose, to our knowledge.

Our approach bridges this gap by integrating AR into curriculum development through a Youth Participatory Action Research (YPAR) \cite{anyon2018systematic} framework, thereby enabling students to actively engage with and critically reflect on linguistic justice issues. This method not only makes the learning process more culturally relevant and personally meaningful to students but also provides them the opportunity to contribute actively to their educational experiences and share experiential knowledge with their peers.

Social justice-related uses of AR remain rare; the closest project to our own that we are aware of is \cite{hidalgo2015augmented}, which describes a new multi-modal media object, Augmented Fotonovela, which uses AR to bring fotonovelas (a traditional print medium) to life by superimposing video interviews over images that are activated and begin streaming in real-time with the use of a free app. Like our project, this work illustrates the use of AR to focus on the lived experiences of Communities of Color and engage them as co-collaborators in a creative research process. However, in \cite{hidalgo2015augmented}, collaborators were adults rather than youth. Our project is therefore unique by involving youth of color in an educational setting.

\section{Method}
\subsection{SKILLS}
\label{skills}
SKILLS (School Kids Investigating Language in Life and Society) \cite{bucholtz2014sociolinguistic, bucholtz2017language, lawson2017sociolinguistic} is a social justice-oriented academic partnership between UC Santa Barbara and local teachers that prepares and motivates California’s public school students, most of them from low-income Latinx families, for higher education by giving them hands-on experience in studying language, race, power, and identity. The main goal of SKILLS is to promote educational justice for students from  underrepresented, marginalized communities to embrace student agency and combat deficit ideologies around their linguistic knowledge. SKILLS puts students at the center of knowledge creation and social change by guiding them through the process of carrying out original research and community action projects on language in their own peer groups, families, and communities. SKILLS embodies a politically engaged approach to education that centers on students' experiences and provides them agency through participatory research and community action. This program not only educates students about linguistic justice but also encourages them to become active participants in social change. By grounding a linguistics curriculum in students' lived experiences, SKILLS connects academic concepts with real-world examples. Instruction is provided using innovative pedagogical methods via interactive mini-lectures supported by video-based examples of the concept at play and follow-up activities such as discussions or creations of artifacts to reflect on and demonstrate students’ understandings of the concepts. AR presents an exciting opportunity to more deeply engage students in building shared and more authentic knowledge and experiences about linguistic justice that goes beyond what conventional paper-pencil and discussion tasks can afford, creating more immediacy and opportunity for joint exploration. Students are also trained in research methods that guide them to conduct an independent research or community action project on the language justice issue of their choice. At the end of the program, students present their project results in a poster session at a special SKILLS Day event at UC Santa Barbara. 

\begin{figure}
    
    \includegraphics[scale=0.6]{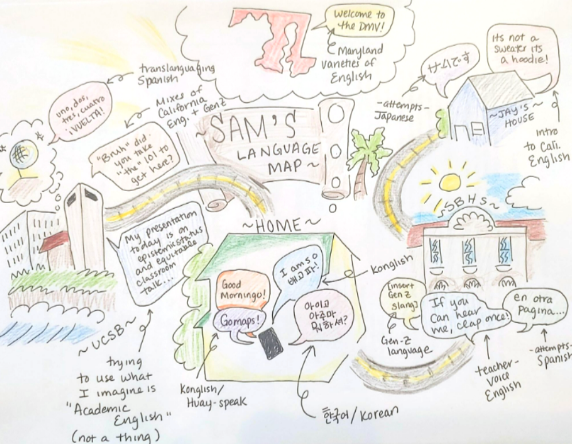}
    \caption{ Example of a Language Map showing the creator’s use of different languages in different spaces.}
    \label{fig:sam_language_map}
\end{figure}

Inspired by linguistic cartography, a popular and effective SKILLS classroom activity is the drawing of a \textbf{language map} which students create as part of a unit on linguistic diversity and linguistic justice. Each student first writes a list of the different physical and virtual places they regularly visit, the people they interact with in each space and the kind of language(s) they use when talking to people in each of these spaces. Figure \ref{fig:sam_language_map} shows an example of a language map. The students then represent this information visually by drawing their language map on chart paper. Students share and discuss their language maps in small groups. This gives them a peek into others’ everyday language experiences and allows them to share their own experiences with others. However, this method of sharing does not provide all the information one needs to fully understand the language map. In order for students to gain direct insight into their peers’ linguistic and social worlds, it would be much more pedagogically powerful for students to create interactive language maps which come alive when the AR device is pointed to the drawn paper map, where students can interact with the characters, hear the actual language being spoken to different people in the various social domains of language use. 
\begin{figure}
    
    \includegraphics[scale=0.6]{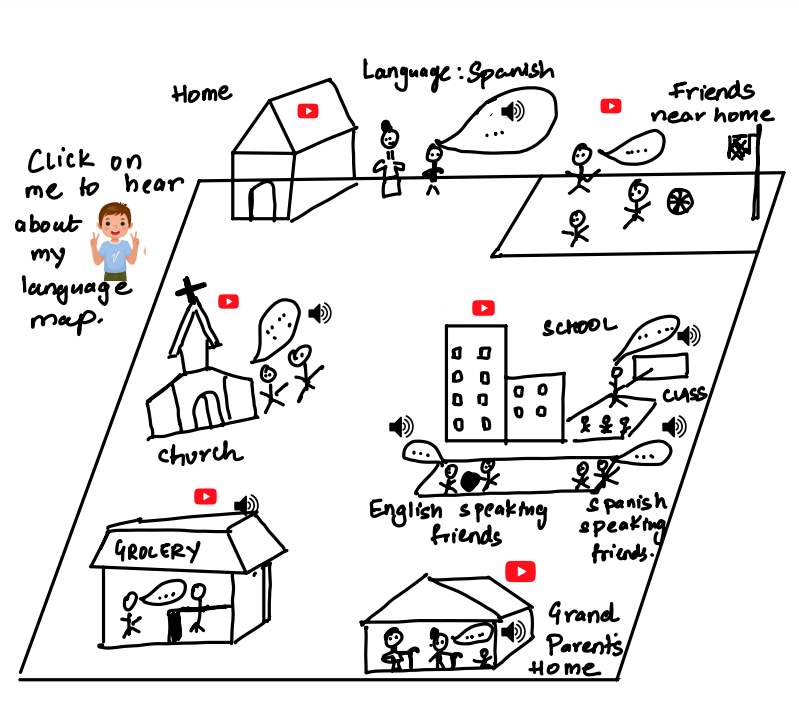}
    \caption{An AR-based Language Map. The user can interact with the characters by clicking on them. The map can include both  speech audio and  video of social interaction within the space.}
    \label{fig:ar_language_map}
\end{figure}

\subsection{AR LANGUAGE MAP: A SPECULATIVE ARTIFACT}
The AR Language Map is a speculative educational tool that is an extension to the SKILLS language map activity explained in section \ref{skills}. It aims to enrich the current language map activity by incorporating AR to offer a dynamic, interactive learning experience. In this enhanced version, students still begin by noting the diverse settings they frequent, the individuals they interact with there, and the languages or language varieties they employ. This information forms the basis of their personalized language maps. Similar to the study in \cite{hidalgo2015augmented}, the AR enhancement transforms these static maps into immersive experiences. Instead of merely viewing flat representations, students can engage with virtual embodiments of their peers' linguistic journeys. Through AR technology, the language maps leap off the page, allowing users to interact with three-dimensional representations of the documented social environments and hear the actual languages and dialects spoken by real people in students’ lives. This direct engagement offers a more profound comprehension of peers' linguistic realities, showcasing the nuances of language use in different social contexts and the emotional resonances attached to these linguistic choices. The AR Language Map shown in figure \ref{fig:ar_language_map} visualizes this speculative interactive artifact. The students write text sentences and create characters when building their AR Language Map. When the user points to the AR Language Map, they will be able to see the 3D view with various spaces and characters whom they can interact with through audio and video recordings. The artifact also includes a narration component wherein the creator of the map has the chance to share the feelings they experience when engaging in each of these spaces and the language choices they make in each context. Touching different parts of the language map plays different pieces of the audio narrative recorded by the student.

Particularly for students who have encountered linguistic racism, a mere visual representation lacks the depth to fully express their experiences and its emotional impacts. The AR Language Map thus serves as a potent educational tool in combating linguistic racism and fostering empathy among students by immersively showcasing the richness and complexity of linguistic diversity. 

We envision the process of building AR Language Maps in the SKILLS classroom using a YPAR approach. First, we first introduce the students to AR technology, its features and its potential using examples of AR use cases. We then engage students in a brainstorming session and guide them to decide how they would like to convert their paper language map into an interactive AR Language Map. The AR component requires an easy-to-use interface for the students to add in their media, characters, and text, build the design, and test their AR Language maps. Finally, the resulting maps are showcased instead of poster presentations at SKILLS Day.

\section{Conclusion}
Building on the conceptual foundation laid by the AR Language Map, future directions will focus on the development of a comprehensive curriculum that employs YPAR methodologies to integrate social justice and linguistic justice concepts into AR-based educational projects. This curriculum aims not only to deepen students' understanding of these critical issues but also to engage their agency as co-creators of their learning experiences. AR is a powerful tool to bring into the SKILLS curriculum to enhance our mission to foster students’ development of agents of linguistic justice. Incorporating AR into the SKILLS curriculum offers a novel and impactful way to teach and understand linguistic justice. Through immersive technology, students can engage with complex concepts in a more meaningful and personal manner, fostering a deeper appreciation for linguistic diversity and the importance of linguistic rights.

The next phase of the project will involve the integration of AI with AR characters, enabling more interactive and responsive dialogues within the AR Language Map. This advancement will further personalize the learning experience, allowing students to engage in simulated conversations that reflect real-world linguistic interactions. Such technology-driven approaches will change the way linguistic justice is taught and understood, making it more accessible, engaging, and impactful for students of all ages.

Through this workshop, we hope to share our ideas in an interdisciplinary space and find potential collaborations to leverage the full potential of AR to challenge and expand the current paradigms of linguistic justice education, ensuring it is comprehensive, inclusive, and transformative.

\bibliographystyle{ACM-Reference-Format}
\bibliography{submission}

\end{CJK}
\end{document}